\documentclass[10pt,conference]{IEEEtran}
\IEEEoverridecommandlockouts

\newcommand{\Lambdaop}{\hat{\Lambda}}
\usepackage[letterpaper]{geometry}
\usepackage{xcolor}
\newif\ifreview
\reviewtrue

\newcommand{\revred}[1]{\ifreview\textcolor{red}{[A: #1]}\fi}
\usepackage{cite}
\usepackage{amsmath,amssymb,amsfonts}
\usepackage{algorithmic}
\usepackage{graphicx}
\usepackage{textcomp}
\usepackage{xcolor}
\def\BibTeX{{\rm B\kern-.05em{\sc i\kern-.025em b}\kern-.08em
    T\kern-.1667em\lower.7ex\hbox{E}\kern-.125emX}}
\begin{document}

\title{Hybrid Predictive Quantum Feedback: Extending Qubit Lifetimes Beyond the
Wiseman-Milburn Limit 
}
\author{%
  \IEEEauthorblockN{%
   Ali Abu\mbox{-}Nada\IEEEauthorrefmark{1}\quad
    Aryan Iliat\IEEEauthorrefmark{2}\quad
    Russell Ceballos\IEEEauthorrefmark{3}\IEEEauthorrefmark{4}}
  
  \IEEEauthorblockA{\IEEEauthorrefmark{1}Sharjah Maritime Academy, Sharjah, United Arab Emirates\\
  Email: ali.abunada@sma.ac.ae}
  
  \IEEEauthorblockA{\IEEEauthorrefmark{2}School of Physics and Applied Physics, Southern Illinois University,\\
  Carbondale, IL 62903, USA\\
  Email: aryan.iliat@siu.edu}
  
  \IEEEauthorblockA{\IEEEauthorrefmark{3}Department of Physical Sciences, Olive-Harvey College, City Colleges of Chicago,\\
  10001 S Woodlawn Ave, Chicago, IL 60628, USA}
  
  \IEEEauthorblockA{\IEEEauthorrefmark{4}QuSTEAM Initiative, 510 Ronalds St., Iowa City, IA 52245, USA \\
  rceballos@qusteam.org}
}

\maketitle

\begin{abstract}
Amplitude damping sets a fundamental limit on qubit lifetimes, while standard Wiseman--Milburn feedback offers only limited improvement due to single-quadrature measurement and feedback delay. We propose a hybrid enhancement that combines an ancilla-assisted coherent feedback loop with lightweight supervised prediction of the homodyne current. A fast-decaying ancilla coherently recovers information from both field quadratures, while prediction compensates hardware latency by phase-aligning the corrective drive. Analytical treatment yields effective decay-rate suppression, and numerical simulations using realistic superconducting-qubit parameters demonstrate a clear performance gain beyond the Wiseman--Milburn limit, achieving multi-fold extension of the effective $T_1$. The approach is modular and compatible with existing feedback hardware, providing a practical route to converting leaked information into time-advanced quantum control.
\end{abstract}

\section{Introduction}\label{sec:sec1}

The loss of energy from a quantum system to its environment, known as
\emph{amplitude damping}, is one of the main causes of decoherence in quantum
technology. When an excited qubit emits a photon and falls to its ground state,
the superposition needed for quantum computation, communication, and sensing is
destroyed \cite{breuer2002theory,nielsen2010quantum}. The timescale of this
process, the relaxation time \(T_1\), therefore places a strict limit on qubit
performance: a larger \(T_1\) means higher-fidelity gates and less overhead in
quantum error correction \cite{Krantz2019APR,Klimov2024NC,Fowler2012PRA}.

Two main approaches are commonly used to extend \(T_1\). The first is
\emph{environment engineering}, which reduces the electromagnetic density of
states at the qubit frequency, since \(T_1^{-1}\propto S_\perp(\omega_0)\).
Examples include Purcell filters
\cite{purcell1946,Reagor2013APL,Heinsoo2018PRAppl,Jeffrey2014PRL,Sete2015PRA,
Spring2025PRXQ} and photonic bandgap structures
\cite{Yablonovitch1987PRL,Lodahl2015RMP}. Although effective, these require
precise fabrication and careful frequency matching. The second approach is
\emph{active control}. Dynamical decoupling (DD) can refocus low-frequency
phase noise
\cite{Cywinski2008PRB,Biercuk2009Nature,Uys2009PRL,Bylander2011NatPhys,
abunada2025dynamicscontrolcoupledquantum,Viola1998PRL,Lidar2003JMO,
Byrd2001PRA,Wu2002PRL}, but offers little protection against amplitude damping,
which is driven by high-frequency fluctuations near \(\omega_0\)
\cite{Ezzell2023PRApplied,Clerk2010RMP,Ithier2005PRB,Green2013NJP,
Bylander2011NatPhys}.

A different idea was introduced by Wiseman and Milburn
\cite{Wiseman1993PRL,WisemanMilburnBook,Doherty1999PRA}. Instead of modifying
the environment or applying predetermined pulses, their framework uses
\emph{continuous measurement and feedback}. The field emitted by the qubit is
monitored in real time, usually by homodyne detection, producing a measurement
signal. This signal is immediately fed back as a control operation. In this
way, the measurement itself becomes part of the dynamics and can slow down the
decay \cite{WANG2001221}.
\begin{figure}[t]
    \centering
    \includegraphics[width=\linewidth,height=2.2in,keepaspectratio]{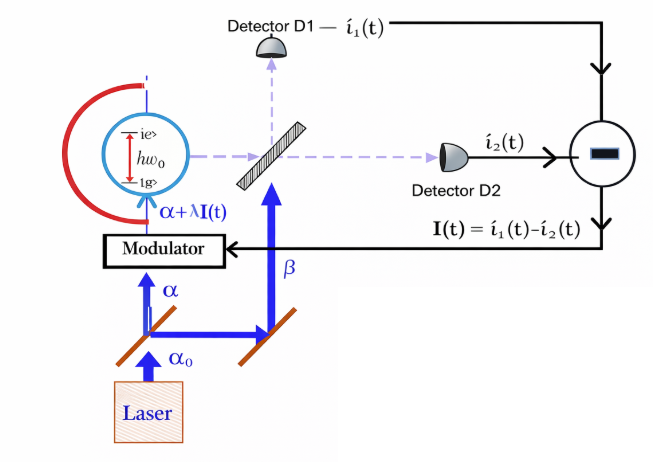}
    \caption{Schematic of the feedback–controlled homodyne interferometer used to monitor and stabilize the emission of a two–level system with transition frequency $\omega_{0}$. A coherent laser drive is split into a probe arm ($\alpha_{0}\!\to\!\alpha$) and a reference arm ($\beta$). The emitter’s scattered field is combined with the reference on a 50:50 beam splitter and measured on two photodetectors (D1, D2), producing photocurrents $i_{1}(t)$ and $i_{2}(t)$. Their difference $I(t)=i_{1}(t)-i_{2}(t)$ is fed back through a modulator to adjust the drive amplitude $\alpha+\lambda I(t)$, enabling real-time stabilization/noise control of the emitter dynamics.}
    \label{fig:fig1}
\end{figure}
Figure~\ref{fig:fig1} shows the basic W-M feedback circuit.
A laser is tuned to the qubit transition frequency \(\omega_0\), the natural
energy splitting between \(|e\rangle\) and \(|g\rangle\). The laser is split
into two paths. One path, labeled \(\beta\), is sent directly to the detectors
and serves as the strong local oscillator (LO) needed for homodyne detection.
The other path, labeled \(\alpha\), drives the qubit. This probe field is later
modulated by the feedback as \(\alpha+\lambda I(t)\). The qubit emits a weak
field at the same frequency \(\omega_0\). This weak emission and the strong LO
interfere on a 50:50 beam splitter, and the two output ports are measured by
detectors D1 and D2, producing photocurrents \(i_1(t)\) and \(i_2(t)\).

To understand what the detectors measure, recall that the qubit’s emitted field
can be written in terms of two independent components, called quadratures, $
\hat{E}_{\mathrm{field}}(t)
= \hat{E}_x(t)\cos(\omega_0 t) + \hat{E}_y(t)\sin(\omega_0 t)$,
where \(\hat{E}_x\) is the in-phase part and \(\hat{E}_y\) is the
out-of-phase part. For a two-level system, these quadratures correspond to the
Pauli operators, \(\hat{E}_x\propto\sigma_x^{(S)}\) and
\(\hat{E}_y\propto\sigma_y^{(S)}\). The LO field
\(E_{\mathrm{LO}}(t)=E_0\cos(\omega_0 t+\phi)\) provides a stable phase
reference. When the LO and emitted field meet at the beam splitter, each
detector measures a large intensity dominated by the LO, so neither
\(i_1(t)\) nor \(i_2(t)\) alone carries clear information about the qubit.

The key step is to subtract the detector outputs,$
I(t)=i_1(t)-i_2(t)$,
which removes the large DC background from the LO and leaves only the small
interference term between the LO and the qubit emission. In the strong-LO
limit, this interference is proportional to a chosen quadrature, $
I(t) \propto E_0\!\big[\hat{E}_x(t)\cos\phi + \hat{E}_y(t)\sin\phi\big]
      \propto \langle\sigma_\phi^{(S)}\rangle$,
where \(\sigma_\phi^{(S)}=\sigma_x^{(S)}\cos\phi+\sigma_y^{(S)}\sin\phi\). Thus,
by selecting the LO phase \(\phi\), the detector measures either
\(\sigma_x\) (in-phase) or \(\sigma_y\) (out-of-phase). This gives a clean,
amplified measurement of one quadrature of the qubit’s emission. The homodyne
current is usually written as $
I(t)
=\sqrt{\eta\gamma}\,\langle\sigma_\phi\rangle_c(t)
 + \frac{\xi(t)}{\sqrt{\eta}}$,
where \(\eta\) is the detector efficiency, \(\gamma\) is the spontaneous
emission rate, \(\langle\sigma_\phi\rangle_c(t)\) is the measured quadrature,
and \(\xi(t)\) is the measurement noise.

This current is fed back to modulate the drive field as \(\alpha+\lambda I(t)\),
so that the correction field interferes with the qubit’s emitted field and
partially cancels it, slowing the decay. However, the W-M method has two fixed limitations that cannot be avoided by
simply improving the hardware. First, homodyne detection measures only
\emph{one} quadrature of the emitted field. This means the feedback knows only
half of what the qubit is doing. It can push the qubit back along one
direction of the Bloch sphere, but the motion along the unmeasured direction is
completely invisible and therefore cannot be corrected. As a result, part of
the spontaneous emission always escapes, and the lifetime improvement is
limited to \(T_1^{\mathrm{eff}}\le 2T_1\). Second, real experiments have a small
but unavoidable electronic delay. Because the qubit evolves very fast, even a
few nanoseconds of delay causes the correction to arrive with the wrong phase,
making it much less effective and sometimes even harmful. These two issues,
missing information and delayed correction, motivate our hybrid predictive
feedback strategy.

\textbf{Step 1: Add an ancilla qubit (to recover the missing quadrature).}
We introduce a second qubit that is coherently coupled to the main system
qubit. The feedback signal \(I(t)\) does not drive the system directly; it
drives the ancilla. As the ancilla precesses under its Hamiltonian, its motion
naturally mixes the \(x\) and \(y\) components of its Bloch vector. Through the
coherent coupling between system and ancilla, this mixed motion acts back on
the system and generates a correction that has \emph{both} quadratures. In this
sense, the ancilla behaves like a quantum ``translator'': it takes the
one-dimensional classical measurement signal and turns it into a full
two-dimensional quantum correction. This allows the feedback to act along both
directions of the Bloch sphere, overcoming the single-quadrature restriction of
the W-M method.
\textbf{Step 2: Predict the future (to fix the delay).}
The second limitation of the W--M scheme is hardware delay: cables and
electronics introduce a small delay \(\tau\), so using the measured current
\(I(t)\) directly means the correction always arrives slightly too late. By the
time the feedback reaches the qubit, the qubit’s state has already changed,
and the correction is applied with the wrong phase. To overcome this, we train
a supervised machine-learning model to predict the future homodyne signal,
\(\widehat{I}(t{+}\tau)\), from the recent history of the measurement record.
As shown in Fig.~\ref{fig:fig4}, the predicted current (blue curve) closely
matches the true future current \(I(t{+}\tau)\) (red curve), while the present
signal \(I(t)\) (dashed gray) is clearly out of sync. By using the predicted
future value instead of the delayed measurement, the feedback arrives
\emph{in phase} with the qubit’s emission, restoring the destructive
interference needed for strong suppression of spontaneous decay.

\begin{figure}[t]
    \centering
    \includegraphics[width=\linewidth,height=2.2in,keepaspectratio]{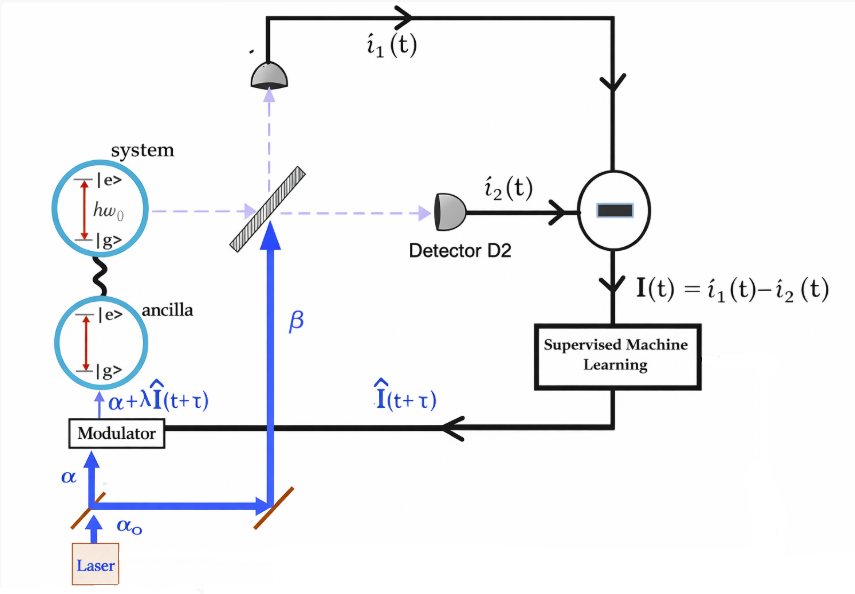}
   \caption{\textbf{Hybrid predictive feedback scheme.} 
The emitted light from the system is measured by two detectors (D1, D2), producing the homodyne current 
$I(t)=i_1(t)-i_2(t)$. 
A supervised machine-learning model predicts the future signal $\widehat{I}(t{+}\tau)$ to overcome the feedback delay~$\tau$. 
The predicted current drives the modulator, which adjusts the laser field 
to $\alpha+\lambda\widehat{I}(t{+}\tau)$. 
This corrected field interacts with a coherently coupled \emph{ancilla} qubit, which then steers the main system qubit. 
Together, the ancilla and the ML predictor allow the feedback to act on both field quadratures and stay in phase with the emission, achieving stronger suppression of spontaneous decay than the standard W-M loop.}
    \label{fig:fig2}
\end{figure}

Figure~\ref{fig:fig2} shows the full hybrid loop. The emitted field is
measured, converted into the homodyne current, and then predicted forward in
time by the ML model. This predicted current drives the modulator, which
updates the probe field. The corrected drive interacts with the ancilla, which
applies a fast, coherent, two-quadrature correction to the system. Together,
the ancilla and the predictor suppress spontaneous emission \emph{as it
happens}, achieving much stronger stabilization than the classical W--M loop.\\
\revred{We emphasize that this work targets energy relaxation ($T_1$)
and amplitude damping. Dephasing noise is largely orthogonal and can be
addressed using established techniques such as echo and dynamical
decoupling. The proposed hybrid predictive loop is compatible with such
methods and can be extended to multi-channel feedback for correlated
noise, which we leave for future work.}

\section{From Classical Feedback to Hybrid Predictive Quantum Control}\label{sec:sec2}
Real-time feedback has long been used to stabilize or cool quantum systems,
but conventional W-M feedback remains limited by two 
fundamental constraints: (i) it acts only on a single field quadrature, so 
half of the available information is lost to the environment, and 
(ii) it operates with a finite delay $\tau$ between measurement and control, 
so the corrective tone always arrives \emph{after} the emission it seeks to cancel. 
These restrictions impose a hard twofold ceiling on the achievable lifetime 
extension ($T_1^{\mathrm{eff}}\!\le2T_1$). 

To overcome these limits, we introduce a hybrid architecture that 
combines a \emph{quantum-coherent ancilla} with a 
\emph{supervised machine-learning predictor}.  
The ancilla couples coherently to the system, allowing the feedback to act 
simultaneously on both field quadratures and thus recover the half of the 
information lost in W-M control.  
Meanwhile, the machine-learning predictor forecasts the future homodyne 
current $\widehat{I}(t{+}\tau)$, enabling phase-aligned actuation that 
compensates the physical propagation delay in the loop.  
Together, these upgrades convert classical measurement-based feedback into a 
fully predictive quantum-control protocol capable of surpassing the W-M 
lifetime limit.  
The following subsections describe these two components in detail.

\subsection{Effective Dynamics of Ancilla-Assisted Feedback}\label{sec:sec2a}

In the original W-M scheme
\cite{Wiseman1993PRL,WisemanMilburnBook,WANG2001221}, 
a decaying qubit emits light that, like any electromagnetic wave, contains two
orthogonal \emph{quadratures}: an in-phase ($x$) component and an out-of-phase
($y$) component. Balanced homodyne detection
(Fig.~\ref{fig:fig1}) mixes this weak field with a strong local oscillator (LO)
of phase~$\phi$ and measures only \emph{one} quadrature at a time. The two
photodetectors produce a single classical signal, the homodyne current, $
I(t)\ \propto\ \big\langle \sigma_{\phi}^{(S)} \big\rangle_c(t),
\quad
\phi=0~(x),\quad \phi=\tfrac{\pi}{2}~(y)$.
Because $I(t)$ contains information from only one quadrature, the feedback loop
can push back only along one direction of the Bloch sphere and therefore cannot
fully cancel spontaneous emission.

\paragraph*{A quantum helper (ancilla) to recover the missing quadrature.}
To overcome this limitation, we introduce a second qubit, the \emph{ancilla}, and
let the measured current drive the ancilla instead of the system
(Fig.~\ref{fig:fig2}). The ancilla is \emph{coherently} coupled to the system,
so the feedback now enters the loop at the quantum level. This allows the loop
to affect \emph{both} quadratures of the system’s motion. Intuitively, the
ancilla acts as a “quantum translator”: even though $I(t)$ carries only
one quadrature, the ancilla’s own coherent evolution mixes its $x$ and $y$
components and converts the one-dimensional classical signal into a phase-aware
two-quadrature correction on the system.

\paragraph*{Conditional stochastic evolution.}
Let $\rho_{SA,c}$ be the conditional state of the system (S) and ancilla (A).
Under continuous homodyne monitoring, it evolves according to
\cite{WisemanMilburnBook,carmichael1993,JacobsSteck2006}
\begin{equation}
\begin{split}
d\rho_{SA,c} &=
 -\frac{i}{\hbar}\!\left[ H_0 + H_{\mathrm{int}},\, \rho_{SA,c}\right] dt
 + \gamma\, \mathcal{D}\!\left[\sigma_-^{(S)}\right]\rho_{SA,c}\, dt \\
&\quad + \sqrt{\eta\gamma}\, \mathcal{H}\!\left[\sigma_-^{(S)}\right]\rho_{SA,c}\, dW(t),
\end{split}
\label{eq:eq1}
\end{equation}
where $\gamma$ is the spontaneous-emission rate, $\eta$ is the detector
efficiency, and $dW(t)$ is a Wiener increment ($\mathbb{E}[dW]=0$,
$dW^2=dt$). The dissipator $\mathcal{D}$ describes decay and
$\mathcal{H}$ describes measurement back-action.

\paragraph*{Hamiltonians and coherent coupling.}
\begin{equation}
\begin{aligned}
H_0 &= H_S + H_A + H_{\mathrm{drive}}^{(A)}(t),\\[2pt]
H_S &= \frac{\hbar\omega_S}{2}\, \sigma_z^{(S)}, \qquad
H_A = \frac{\hbar\omega_A}{2}\, \sigma_z^{(A)},\\[2pt]
H_{\mathrm{int}} &= \hbar g \!\left( \sigma_x^{(S)}\sigma_x^{(A)}
                 + \sigma_y^{(S)}\sigma_y^{(A)} \right),
\end{aligned}
\label{eq:eq2}
\end{equation}
where $H_S$ and $H_A$ are the system and ancilla Hamiltonians, 
$\omega_S$ and $\omega_A$ are their transition frequencies, and $g$ is the
coherent coupling rate.

\paragraph*{Feedback acts on the ancilla.}
The feedback signal controls the ancilla:
\begin{equation}
H_{\mathrm{drive}}^{(A)}(t)
 = \hbar\, u(t)\, \sigma_q^{(A)},\qquad
u(t) = \lambda\, I(t),\quad q\in\{x,y\},
\label{eq:eq3}
\end{equation}
with gain $\lambda$ and axis $q$ set by the LO phase. The homodyne current takes
the explicit form
\begin{equation}
  I(t) = \sqrt{\eta \gamma}\,\langle \sigma_{\phi}^{(S)} \rangle_c(t)
         + \frac{\xi(t)}{\sqrt{\eta}},
\label{eq:eq4}
\end{equation}
where $\xi(t)$ is white shot noise. Thus the ancilla is driven by a signal that
originates entirely from the system’s emission.

\paragraph*{How one quadrature becomes two.}
Although the drive $u(t)$ pushes the ancilla along one axis (say, $x$),
the ancilla precesses under its Hamiltonian $H_A=(\hbar\omega_A/2)\sigma_z^{(A)}$.
This rotation mixes its $x$ and $y$ components:
\begin{equation}
  \sigma_x^{(A)}(t)
  = \cos(\omega_A t)\,\sigma_x^{(A)}
  + \sin(\omega_A t)\,\sigma_y^{(A)}.
  \label{eq:eq5}
\end{equation}
Through $H_{\mathrm{int}}$, these mixed quadratures couple back into the system’s
$\sigma_x^{(S)}$ and $\sigma_y^{(S)}$ operators, effectively generating a
two-quadrature correction. The ancilla therefore acts as a \emph{quantum
converter}: it transforms a single-number classical signal into a coherent,
phase-aware control action that addresses both quadratures.

\paragraph*{Averaged (deterministic) dynamics.}
Averaging Eq.~\eqref{eq:eq1} over the measurement noise gives the unconditional
Lindblad evolution
\begin{equation}
\begin{aligned}
\dot{\rho}_{SA}
&= -\frac{i}{\hbar}
   \left[ H_0 + H_{\mathrm{int}}
   + \frac{\hbar}{2}\big(c^\dagger \Lambdaop + \Lambdaop c\big),
   \rho_{SA} \right]\\
&\quad + \mathcal{D}[c - i\Lambdaop]\rho_{SA}
       + (1-\eta)\,\mathcal{D}[\Lambdaop]\rho_{SA},
\end{aligned}
\label{eq:eq6}
\end{equation}
where $c=\sqrt{\gamma}\sigma_-^{(S)}$ is the system emission operator and
$\Lambdaop=\lambda\sigma_q^{(A)}$ is the ancilla feedback operator.
The coherent term produces Hamiltonian shifts due to feedback, while the
dissipators show explicitly how the ancilla’s radiation destructively
interferes with the system’s emission, reducing the effective decay rate.
Setting $\Lambdaop=\lambda\sigma_q^{(S)}$ recovers the classical W--M master
equation~\cite{WANG2001221}, confirming that our approach is a
\emph{quantum-coherent generalization} of W--M feedback.

In Sec.~\ref{sec:sec2b} we extend this model by driving the ancilla with a
short-horizon \emph{prediction} of the homodyne current rather than the measured
value itself, $u(t)=\lambda\,\widehat{I}(t+\tau)$, where
$\widehat{I}(t+\tau)$ is a supervised ML estimate of the future signal. This
keeps the feedback phase-aligned despite electronic delay and further
strengthens the suppression of spontaneous emission.

\subsection{Supervised Machine-Learning Prediction of the Homodyne Current}
\label{sec:sec2b}

In the feedback setup, the balanced homodyne detector continuously measures the quadrature
of the emitted field and produces an analog voltage signal \(I(t)\).
This signal fluctuates because of quantum noise, yet it carries valuable information about
the instantaneous system state.
In any real experimental setup, there exists a finite electronic latency \(\tau\),
arising from cables, filters, and data-processing circuits.
As a result, a control signal calculated at time \(t\) will only reach the ancilla
and influence the system at time \(t{+}\tau\).
If feedback were based solely on the instantaneous value \(I(t)\),
it would arrive too late to counteract the system’s evolution.
To overcome this, we train a supervised machine-learning (ML) model to
\emph{predict the future current} \(I(t{+}\tau)\)
from the recent measurement history.
The predicted signal is then used to drive the ancilla with a phase-leading control tone,
so that, after the loop delay, it arrives synchronized with the actual optical field.

The experiment proceeds in two distinct stages: a \emph{training stage} and a \emph{prediction stage}.

\begin{figure}[t]
    \centering
    \includegraphics[width=\linewidth,height=2.2in,keepaspectratio]{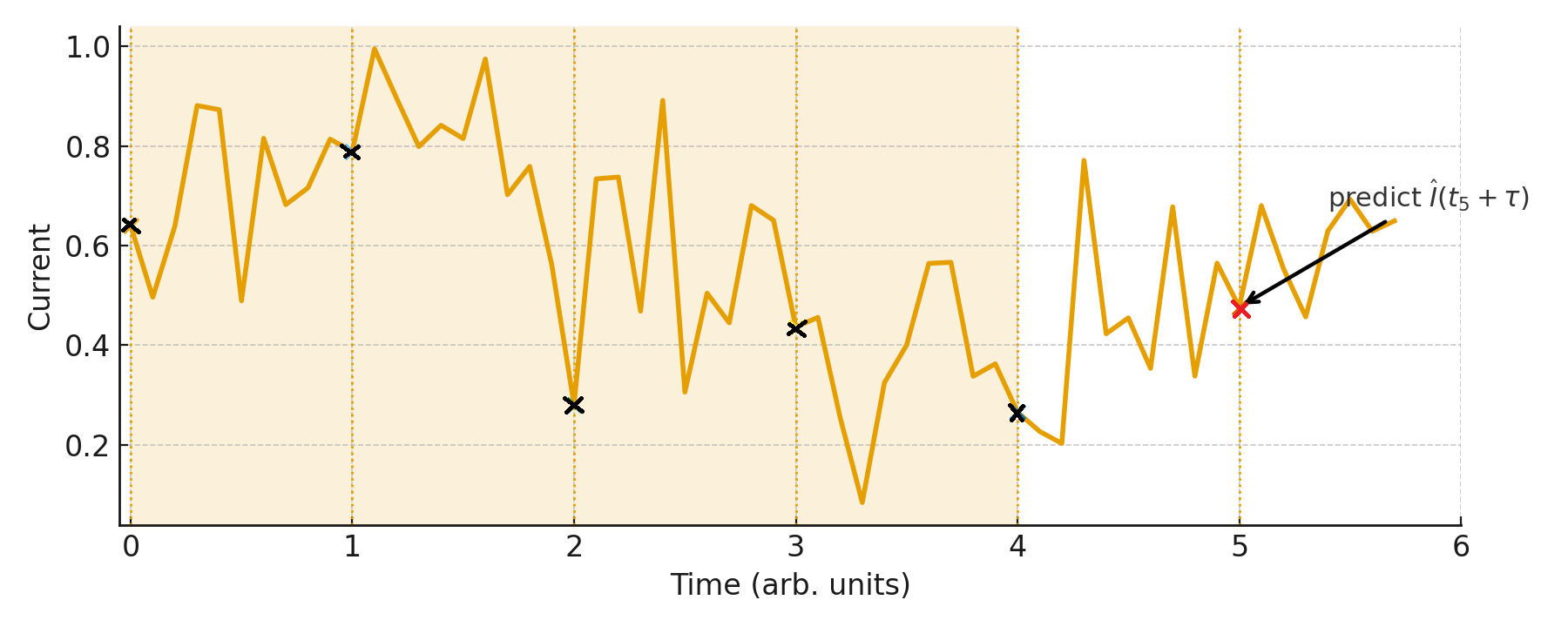}
    \caption{\textbf{Digitizing the homodyne signal and forming training samples.}
    The orange curve shows the measured homodyne current \(I(t)\) after analog-to-digital conversion.
    Each black cross marks one of the five most recent samples 
    \([I(t_{i-5}),\ldots,I(t_{i-1})]\) that form the input vector \(\mathbf{x}_i\),
    while the red point represents the next sample \(I(t_i)\), which the network learns to predict.
    This sliding-window process converts the continuous signal into overlapping
    input–output pairs suitable for supervised learning.}
    \label{fig:fig3}
\end{figure}

\begin{figure}[t]
    \centering
    \includegraphics[width=\linewidth,height=2.2in,keepaspectratio]{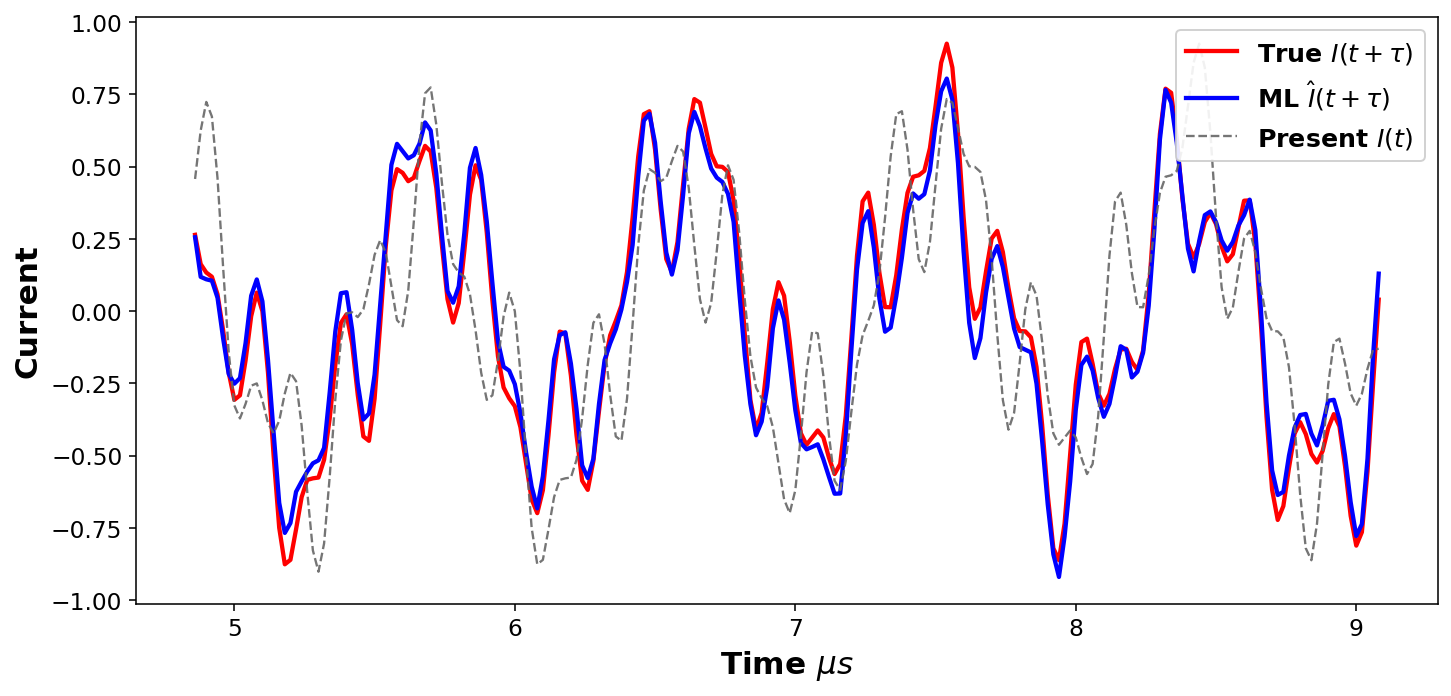}
    \caption{\textbf{Prediction versus measured delayed current.}
    The red curve shows the measured delayed current \(I(t{+}\tau)\),
    while the blue curve shows the ML prediction \(\widehat{I}(t{+}\tau)\)
    obtained from the input windows in Table~\ref{table1}.
    The dashed gray curve indicates the present current \(I(t)\).
    The close overlap between red and blue confirms that the trained network
    accurately anticipates the future signal, allowing the feedback to stay synchronized
    despite the hardware latency.}
    \label{fig:fig4}
\end{figure}

During {Stage 1 (Training)}, the ancilla feedback path is temporarily disconnected,
and a probe device is placed at the ancilla input port to record the homodyne signal
\emph{after the loop delay}.
This produces a delayed record
\(\{I(t_0{+}\tau), I(t_1{+}\tau), I(t_2{+}\tau), \ldots\}\)
sampled at regular intervals \(\Delta t=\tau\),
so that consecutive times differ by one loop delay: \(t_{i+1}-t_i=\tau\).
From this delayed record, the training dataset is constructed by sliding a window
of width \(W=5\), as shown in Fig~\ref{fig:fig3} and  summarized in Table~\ref{table1}.

Each training example consists of five consecutive delayed samples as the input window
and the next delayed sample as the target:
\begin{align}
\mathbf{x}_i &= [I(t_{i-5}{+}\tau),\,I(t_{i-4}{+}\tau),\,I(t_{i-3}{+}\tau),\,I(t_{i-2}{+}\tau),\notag \\ &\,I(t_{i-1}{+}\tau)], \notag\\
y_i &= I(t_i{+}\tau).
\label{eq:eq7}
\end{align}
\begin{table*}[!t]
\centering
\renewcommand{\arraystretch}{1.05}
\setlength{\tabcolsep}{6pt}
\small
\caption{Training pairs constructed from the delayed record. The sampling step is fixed at $\Delta t = \tau$, so each entry in the input window is exactly one delay interval apart.}
\label{table1}
\begin{tabular}{cc}
\hline
\textbf{Input window $\mathbf{x}_i$} & \textbf{Target $y_i$} \\
\hline
$[I(t_0+\tau),\,I(t_1+\tau),\,I(t_2+\tau),\,I(t_3+\tau),\,I(t_4+\tau)]$ & $I(t_5+\tau)$ \\
$[I(t_1+\tau),\,I(t_2+\tau),\,I(t_3+\tau),\,I(t_4+\tau),\,I(t_5+\tau)]$ & $I(t_6+\tau)$ \\
$[I(t_2+\tau),\,I(t_3+\tau),\,I(t_4+\tau),\,I(t_5+\tau),\,I(t_6+\tau)]$ & $I(t_7+\tau)$ \\
$\vdots$ & $\vdots$ \\
\hline
\end{tabular}
\end{table*}

When the network is trained properly, its predicted value for a representative window,
such as the first row of Table~\ref{table1},
satisfies \(\widehat{I}(t_5{+}\tau)\approx I(t_5{+}\tau)\),
and the mean squared error (MSE) over all samples becomes sufficiently small.
As will be explained in the next subsection, this MSE serves as the quantitative measure
of training success.
\emph{Once the model has been trained to predict the next delayed sample accurately,
the probe is removed and the ancilla is reconnected} to begin live feedback operation.

During {Stage 2 (Live prediction)}, the ML continuously receives
the most recent five delayed samples
\([I(t_{i-5}{+}\tau),\ldots,I(t_{i-1}{+}\tau)]\),
feeds them to the trained model,
and obtains a real-time prediction for the next delayed value \(\widehat{I}(t_i{+}\tau)\).
This forecast is used immediately to generate the control drive $u(t)=\lambda\,\widehat{I}(t{+}\tau)$,
so that, after the delay \(\tau\), the control signal arrives phase-aligned
with the true homodyne current.

After we train the network offline on the recorded delayed data, we run it online in real time: at each step it predicts \(\widehat{I}(t{+}\tau)\) from the latest five samples and the controller applies \(u(t)=\lambda\,\widehat{I}(t{+}\tau)\). This turns a delayed, reactive loop into a predictive, anticipatory one.\\
\revred{To isolate the role of learning, we compared the neural predictor
to lightweight linear baselines (moving-average and AR$(5)$ models) with
the same input window. While these heuristics partially compensate delay,
they cannot capture the nonlinear correlations induced by measurement
backaction and coherent system--ancilla dynamics. The neural predictor
achieves the highest correlation with the delayed signal and therefore
the strongest decay suppression at identical latency.}

\subsection{Neural Network Architecture}
\label{sec:sec2c}

\begin{figure}[t]
\centering
\includegraphics[width=1.0\linewidth]{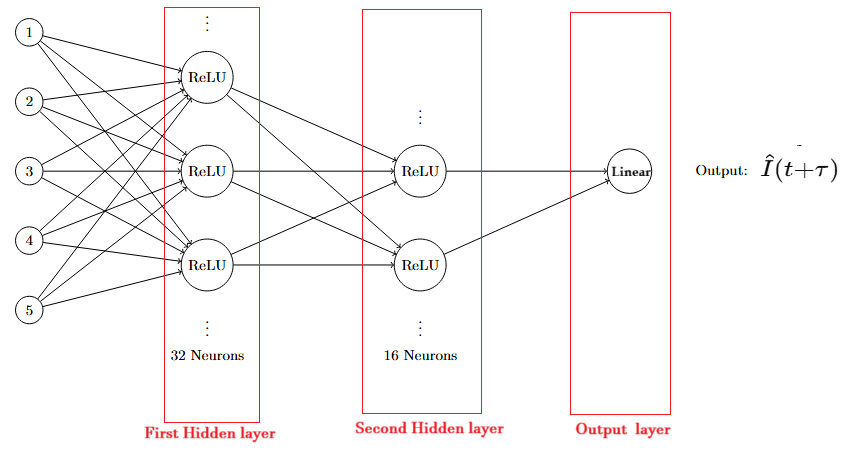}
\caption{\textbf{Neural network used for homodyne-current prediction.}
The input layer receives the last \(W{=}5\) delayed samples
\([I(t_{i-5}{+}\tau),\ldots,I(t_{i-1}{+}\tau)]\).
Two hidden layers (32 and 16 neurons) with ReLU activations
capture nonlinear dependencies in the time series,
and a linear output neuron provides the forecast
\(\widehat{I}(t_i{+}\tau)\) used by the controller.}
\label{fig:fig5}
\end{figure}

The neural network designed for predicting the homodyne current employs a supervised-learning framework similar to that presented in our earlier work~\cite{abunada2025supervisedmachinelearningpredicting}, but is trained here on delayed current samples. It consists of an input layer, two hidden layers, and one output layer.
Together, these layers allow the network to learn the nonlinear relationship
between a short sequence of recent current samples and the next delayed value
\(I(t_i{+}\tau)\).
The hidden layers perform nonlinear transformations that reveal temporal correlations
which a simple linear model cannot capture.
Figure~\ref{fig:fig5} shows the complete architecture.
Each layer is fully connected, every neuron in one layer is connected to every neuron in the next.

\begin{itemize}

\item \textbf{First hidden layer:}
This layer contains 32 neurons.
Each neuron receives all five input features from the sliding window of delayed currents
\([I(t_{i-5}{+}\tau),\ldots,I(t_{i-1}{+}\tau)]\).
Each neuron computes a weighted sum of these inputs, adds a bias,
and applies a Rectified Linear Unit (ReLU) activation~\cite{nair2010relu}:
\begin{equation}
h^{(1)}_j = \text{ReLU}\!\Big(\sum_{k=1}^{5} w^{(1)}_{jk} x_k + b^{(1)}_j\Big),
\label{eq:eq8}
\end{equation}
where \(x_k\) is the \(k\)th input feature (current sample),
\(w^{(1)}_{jk}\) is the weight connecting feature \(k\) to neuron \(j\),
and \(b^{(1)}_j\) is the bias of that neuron.
The ReLU function,
\[
\text{ReLU}(z) =
\begin{cases}
z, & z>0,\\[2pt]
0, & z\le0,
\end{cases}
\]
acts as a nonlinear gate: if the total input \(z\) is positive, the neuron fires
and passes the value forward; otherwise it outputs zero.
This activation allows the network to learn complex current patterns efficiently
without introducing unnecessary computational cost.

\item \textbf{Second hidden layer:}
This layer has 16 neurons, each fully connected to all 32 outputs from the first layer.
The operation is again a weighted sum followed by a ReLU activation:
\begin{equation}
h^{(2)}_j = \text{ReLU}\!\Big(\sum_{k=1}^{32} w^{(2)}_{jk} h^{(1)}_k + b^{(2)}_j\Big),
\label{eq:eq9}
\end{equation}
where \(h^{(1)}_k\) is the activation from neuron \(k\) in the previous layer.
This second layer builds higher-level representations that encode correlations
between temporal features of the current signal.

\item \textbf{Output layer:}
The final layer consists of a single linear neuron that combines all 16 outputs from
the second hidden layer to produce the predicted delayed current:
\begin{equation}
\widehat{y}_i = \sum_{j=1}^{16} w^{(3)}_j h^{(2)}_j + b^{(3)}
\quad\Rightarrow\quad
\widehat{y}_i = \widehat{I}(t_i{+}\tau).
\label{eq:eq10}
\end{equation}
A linear output (no activation) is used because the homodyne current is a real,
unbounded quantity that can take both positive and negative values.
\end{itemize}

Each neuron therefore performs a simple yet fundamental operation:
it takes a weighted sum of its inputs, adds a bias, applies a nonlinear activation
(if applicable), and passes the result forward to the next layer.
Through many such operations across layers, the network gradually learns
to represent the dynamics of the current signal.

\paragraph*{Training objective (MSE loss).}
The supervised learning process adjusts all network parameters
\(\theta=\{W,b\}\) to minimize the mean squared error (MSE)
between the predicted and true delayed currents:
\begin{equation}
\mathcal{L}_{\mathrm{MSE}}(\theta)
= \frac{1}{N}\sum_{i=1}^{N}\big[\widehat{I}(t_i{+}\tau)-I(t_i{+}\tau)\big]^2.
\label{eq:eq11}
\end{equation}
This loss penalizes large deviations and ensures that, on average,
the predicted current closely follows the measured delayed current.
Training uses the Adam optimizer~\cite{Kingma2014adam}
with mini-batches, standardized inputs, and early stopping based on validation MSE
to prevent overfitting.

\paragraph*{Training and prediction phases.}
During the \textbf{training phase}, 50\% of the delayed dataset is used,
where both the inputs \(\mathbf{x}_i\) and the true targets \(y_i\) are known.
The network learns by iteratively adjusting its weights to minimize the MSE loss.
After convergence—when, for instance, \(\widehat{I}(t_5{+}\tau)\approx I(t_5{+}\tau)\) as in Table~\ref{table1},
the model parameters are frozen.

In the \textbf{prediction phase}, the remaining 50\% of the delayed record
is used for testing and live feedback.
Here, the trained network receives only the latest five delayed samples
and outputs the forecast \(\widehat{I}(t_i{+}\tau)\),
without any access to the true \(I(t_i{+}\tau)\) value.
The controller then applies the predicted current as $u(t)=\lambda\,\widehat{I}(t{+}\tau)$,
providing an anticipatory control tone that compensates for the hardware delay.
This two-stage process—offline training and online prediction—transforms the
feedback loop from reactive to predictive. \\
Once trained, the neural network effectively acts as a fast, data-driven predictor:
it continuously processes new measurements, infers the near-future homodyne current,
and drives the feedback controller in advance, thus neutralizing the inherent latency
of the experimental loop.

\section{Estimation of Effective Lifetime and Decay Rate}
\label{sec:sec3}

In this section, we describe how the \emph{effective lifetime} and corresponding \emph{decay rate} of the system are extracted under different feedback configurations.  
Our goal is to quantify how each control strategy, ranging from passive decay to classical, coherent, and predictive feedback, modifies the relaxation dynamics of a two-level system (qubit).  
The fundamental observable used for this comparison is the excited–state population,
\begin{equation}
P_e(t) \equiv \operatorname{Tr}\!\big[\rho(t)\,\Pi_e\big],
\qquad 
\Pi_e = \lvert e\rangle\langle e\rvert,
\label{eq:eq12}
\end{equation}
where \(\rho(t)\) is the system density matrix and \(\Pi_e\) projects onto the excited state \(|e\rangle\).  
The trace operation \(\operatorname{Tr}[\cdot]\) ensures that we extract the total probability of finding the system excited at time \(t\).

For open quantum systems subject to relaxation, the excited-state population typically follows an exponential decay law, 
\begin{equation}
P_e(t) = P_e(0)\, e^{-\Gamma_{\mathrm{eff}} t}, 
\qquad 
T_1^{\mathrm{eff}} = \Gamma_{\mathrm{eff}}^{-1},    
\label{eq:eq13}
\end{equation}
where \(\Gamma_{\mathrm{eff}}\) is the \emph{effective decay rate} and \(T_1^{\mathrm{eff}}\) is the corresponding \emph{effective lifetime}.  
Eq.~\ref{eq:eq13} defines the performance metrics that we later extract from both numerical simulations and the analytical model. 
Physically, the smaller \(\Gamma_{\mathrm{eff}}\) (or equivalently, the larger \(T_1^{\mathrm{eff}}\)), the more effectively the feedback slows the spontaneous emission process.

For a two-level atom or qubit, it is often convenient to use the Pauli-operator basis, 
\(\sigma_- = |g\rangle\langle e|\),
\(\sigma_+ = |e\rangle\langle g|\),
and 
\(\sigma_z = |e\rangle\langle e| - |g\rangle\langle g|\).  
In this notation, the standard Lindblad dissipator for any operator \(L\) is
\begin{equation}
\mathcal{D}[L]\rho = L\rho L^\dagger - \tfrac{1}{2}\{L^\dagger L,\rho\},
\label{eq:eq14}
\end{equation}
where the anticommutator \(\{A,B\}=AB+BA\) guarantees the complete-positivity and trace preservation of the master equation~\cite{breuer2002theory, WisemanMilburnBook}.

\subsection{No Feedback (Natural Decay)}
\label{sec:sec3a}

We begin with the baseline case of spontaneous emission in the absence of any feedback control.  
In the rotating frame and assuming no external drive, the master equation describing energy relaxation at rate \(\gamma\) reads
\begin{equation}
\dot\rho = \gamma\,\mathcal{D}[\sigma_-]\rho.
\label{eq:eq15}
\end{equation}
This equation states that population leaks irreversibly from the excited to the ground state through the jump operator \(\sigma_-\).  To find how the excited-state population evolves, we differentiate Eq.~\eqref{eq:eq12} with respect to time and apply the master equation~\eqref{eq:eq15}:
\begin{equation}
\dot P_e(t) = \operatorname{Tr}\!\big[\dot\rho(t)\,\Pi_e\big]
= \gamma\,\operatorname{Tr}\!\big[\mathcal{D}[\sigma_-]\rho(t)\,\Pi_e\big].
\label{eq:eq16}
\end{equation}
Using standard trace identities (see Appendix~\ref{appendixa}), one obtains
\begin{equation}
\dot P_e(t) = -\gamma\,P_e(t),
\label{eq:eq17}
\end{equation}
whose solution is
\begin{equation}
P_e(t) = P_e(0)e^{-\gamma t},
\quad 
\Gamma_{\mathrm{eff}} = \gamma,
\quad 
T_1 = \gamma^{-1}.
\label{eq:eq18}
\end{equation}

Equation~\eqref{eq:eq18} thus represents the natural (uncontrolled) exponential decay of a two-level system’s excitation, with a relaxation time \(T_1\) equal to the inverse of the spontaneous-emission rate \(\gamma\).

\subsection{Classical (Wiseman–Milburn) Feedback}
\label{sec:sec3b}

We next consider the classical continuous-measurement feedback scheme introduced by
Wiseman and Milburn~\cite{Wiseman1993PRL, WisemanMilburnBook}.
In this setup, a decaying two-level atom emits a fluorescence field that is continuously
monitored through homodyne detection.
The output field carries information about the system’s state, which is converted into
a photocurrent $I(t)$.
This current is then \emph{fed back} to the atom in real time through a control Hamiltonian,
with the aim of stabilizing the emission or slowing spontaneous decay.
In the ideal Markovian limit, where the loop delay is negligible compared to the system
timescale, the unconditional evolution of the atom’s density matrix $\rho(t)$ is governed by
the feedback master equation
\begin{equation}
\dot\rho
= \mathcal{D}\!\big[\sqrt{\gamma}\,\sigma_- - i\lambda\,\sigma_y\big]\rho,
\label{eq:eq19}
\end{equation}
where $\gamma$ is the spontaneous–emission rate,
$\lambda$ is the feedback gain, and
$\sigma_y$ is the Pauli operator defining the quadrature on which feedback acts.
The Lindblad superoperator $\mathcal{D}[L]\rho = L\rho L^\dagger - \tfrac12\{L^\dagger L,\rho\}$
captures the irreversible damping caused by photon emission.
The additional term $-i\lambda\sigma_y$ introduces a Hamiltonian correction proportional
to the instantaneous measurement record, effectively applying a phase–aligned field
that counteracts the decay.\\
For a two-level system, the excited–state population is directly related to the
expectation value of the Pauli operator $\sigma_z$ through $
P_e(t)=\operatorname{Tr}[\rho(t)\Pi_e]
=\tfrac{1}{2}\big(1+\langle\sigma_z\rangle_t\big)$,
where $\langle\sigma_z\rangle_t=\operatorname{Tr}[\rho(t)\sigma_z]$.
Differentiating gives $
\dot P_e(t)=\tfrac{1}{2}\,\tfrac{d}{dt}\langle\sigma_z\rangle_t$.
Hence, solving for $\tfrac{d}{dt}\langle\sigma_z\rangle$ immediately determines
how the excited–state population $P_e(t)$ evolves with time.\\
\paragraph*{Population equation and effective decay rate.}
Carrying out this derivation (see Appendix~\ref{appendixb}) gives
\begin{equation}
\frac{d}{dt}\langle\sigma_z\rangle
= -\,\Gamma(\lambda)\,\big(\langle\sigma_z\rangle+1\big),
\qquad
\Gamma(\lambda)=\gamma - 2\sqrt{{\eta}\gamma}\lambda + 2\lambda^2.
\label{eq:eq20}
\end{equation}
Equation~\eqref{eq:eq20} clearly shows how feedback modifies the natural decay rate.
The first term, $\gamma$, represents the intrinsic spontaneous emission.
The middle term, $-2\sqrt{\gamma}\lambda$, arises from the coherent
interference between the emitted field and the applied feedback field:
a properly phased feedback current can partially cancel the emission amplitude,
reducing the net decay.
The last term, $+2\lambda^2$, corresponds to additional decoherence introduced
by imperfect or noisy feedback—an unavoidable “price” of driving the atom with
a fluctuating current.

\vspace{6pt}
\paragraph*{Detection efficiency and optimal feedback gain.}
In realistic experiments, detectors capture only a fraction $\eta$ of the emitted
light.
Wiseman and Milburn showed that the feedback strength must therefore be limited
to the fraction of information actually measured~\cite{Wiseman1993PRL, WisemanMilburnBook}:
\begin{equation}
\lambda = \tfrac{1}{2}\sqrt{\eta\,\gamma}.
\label{eq:eq21}
\end{equation}
Substituting this optimal value into Eq.~\eqref{eq:eq20}
yields the effective decay rate
\begin{equation}
\Gamma_{\mathrm{WM}} = \gamma\!\left(1 - \tfrac{\eta}{2}\right),
\label{eq:eq22}
\end{equation}
and thus the excited–state population follows
\begin{equation}
P_e(t)=P_e(0)e^{-\Gamma_{\mathrm{WM}}t},
\qquad
T_1^{\mathrm{WM}}=\frac{1}{\gamma(1-\eta/2)}.
\label{eq:eq23}
\end{equation}

\vspace{6pt}

Eq.~\eqref{eq:eq22} reveals that the feedback reduces the effective decay rate
in direct proportion to the measurement efficiency $\eta$.
If all the emitted photons are detected ($\eta=1$),
the feedback current carries complete information about the emission process,
and the decay rate is halved, $\Gamma_{\mathrm{WM}}=\gamma/2$.
Consequently, the lifetime doubles, $T_1^{\mathrm{WM}}=2/\gamma$,
which is the theoretical limit for any purely classical (measurement-based) feedback.
If only half of the emission is captured ($\eta=0.5$),
the lifetime enhancement is modest ($T_1^{\mathrm{WM}}\approx1.33/\gamma$).
This trade-off emphasizes that the power of classical feedback is fundamentally bounded
by how much information can be extracted from the environment:
measurement improves control, but incomplete detection leaves residual decoherence.
\noindent
In summary, by working in terms of $\frac{d}{dt}\langle\sigma_z\rangle$, we obtain
a closed, analytically solvable equation that clearly exposes how feedback gain $\lambda$
and detection efficiency $\eta$ renormalize the decay rate.
The exponential form of $P_e(t)$ in Eq.~\eqref{eq:eq23} provides a convenient way
to extract the effective lifetime $T_1^{\mathrm{WM}}$ directly from simulation or experiment.

\subsection{Ancilla-Assisted and Machine-Learning–Enhanced Feedback}
\label{subsec:ancilla_ML}

While the original W-M feedback loop acts on a single optical quadrature of the emitted field, the next two strategies, \emph{ancilla-assisted} and \emph{machine-learning–enhanced} feedback—introduce additional dynamical resources that further suppress dissipation and extend qubit lifetimes.

\paragraph*{(a) Ancilla-Assisted Feedback.}
In this scheme, the measured homodyne signal is coherently coupled to a secondary two-level system (the \emph{ancilla}) that exchanges excitations with the main qubit. 
The interaction Hamiltonian is
\begin{equation}
H_{\mathrm{int}} = \hbar g\big(\sigma_+^{(S)}\sigma_-^{(A)} + \sigma_-^{(S)}\sigma_+^{(A)}\big),
\label{eq:eq24}
\end{equation}
where,  \(S\) and \(A\) denote the \emph{system} and \emph{ancilla} qubits, respectively, \(\sigma_\pm^{(S/A)}\) are the raising and lowering operators of each qubit, \(g\) is the coherent coupling strength between the two qubits, \(\gamma\) is the intrinsic decay rate of the system qubit.

The ancilla acts as a temporary energy reservoir, coherently exchanging excitations with the system before dissipating them into the environment. 
When the ancilla relaxes much faster than it couples (\(\kappa \gg g\)), where \(\kappa\) is the ancilla’s spontaneous relaxation rate, it can be \emph{adiabatically eliminated}, yielding an effective decay rate for the system~\cite{WisemanMilburnBook,Gough2010coherent}:
\begin{equation}
\Gamma_{\mathrm{anc}} = \frac{\gamma}{1 + C},
\qquad
C = \frac{4g^2}{\kappa\gamma},
\label{eq:eq25}
\end{equation}
where \(C\) is the \emph{cooperativity parameter}, quantifying how strongly the system–ancilla coupling competes with their dissipation channels.  
A higher cooperativity \(C\) results in a slower decay, leading to an extended relaxation time
\begin{equation}
T_1^{(\mathrm{anc})} = \frac{1 + C}{\gamma}.
\label{eq:eq26}
\end{equation}
The ancilla-assisted improvement arises from coherent feedback interference,
not from adding an extra dissipative reservoir; hence the effective decay is
suppressed by the cooperativity factor $C$.

\paragraph*{(b) Machine-Learning–Enhanced Feedback (Prediction Only).}
To mitigate the hardware latency \(\tau\) present in real feedback loops, we introduce a trained neural network that predicts the \emph{future} homodyne current a short time \(\tau\) ahead.  
The control signal applied to the ancilla is therefore based on this forecast:
\begin{equation}
u_{\mathrm{ML}}(t) = \lambda\,\widehat{I}(t{+}\tau),
\label{eq:eq27}
\end{equation}
where, \(\lambda\) is the scalar feedback gain, \(I(t)\) is the measured (true) homodyne current at time \(t\), \(\widehat{I}(t{+}\tau)\) is the predicted current for time \(t+\tau\), obtained from the most recent delayed samples (Sec.~\ref{sec:sec2b}), \(\tau\) is the loop delay arising from analog digital conversion, filtering, and signal routing.

The prediction quality is quantified by the correlation coefficient
\begin{equation}
r = \mathrm{corr}\!\big(\widehat{I}(t{+}\tau),\, I(t{+}\tau)\big),
\qquad 0 \le r \le 1,
\label{eq:eq28}
\end{equation}
where \(r=1\) represents perfect anticipation of the true delayed signal, while smaller \(r\) indicates poorer predictive alignment.  
The residual, uncorrelated noise fraction scales as \((1 - r^2)\), which renormalizes the effective decay rate from Eq.~\eqref{eq:eq25}:
\begin{equation}
\Gamma_{\mathrm{ML}}
= \Gamma_{\mathrm{anc}}(1 - r^2)
= \frac{\gamma}{1 + C}(1 - r^2).
\label{eq:eq29}
\end{equation}
The corresponding lifetime enhancement becomes
\begin{equation}
T_1^{(\mathrm{ML})}
= \frac{1 + C}{\gamma(1 - r^2)}.
\label{eq:eq30}
\end{equation}

Equation~\eqref{eq:eq30} shows that both the coherent coupling (\(C\)) and the predictive accuracy (\(r\)) act multiplicatively to extend the qubit lifetime.  
In the limit of ideal prediction (\(r\to 1\)) and strong coupling (\(C\gg1\)), the effective relaxation time can greatly exceed the natural \(T_1 = 1/\gamma\).

For a complete derivation of the effective decay rates \(\Gamma_{\mathrm{anc}}\) and \(\Gamma_{\mathrm{ML}}\) starting from the full system–ancilla master equation, see Appendix~\ref{appendixc}.

\section{Results and Discussion}
\label{sec:sec4}
\revred{We agree that lifetime extension alone does not imply fault
tolerance. However, increasing $T_1$ directly reduces the amplitude-damping
error per gate and per error-correction cycle, thereby improving average
gate fidelity and lowering QEC overhead. Our protocol should be viewed
as a control-layer primitive that complements calibration, optimal
control, and quantum error correction rather than replacing them.}

\begin{figure}[t]
    \centering
    \includegraphics[width=1.0\linewidth]{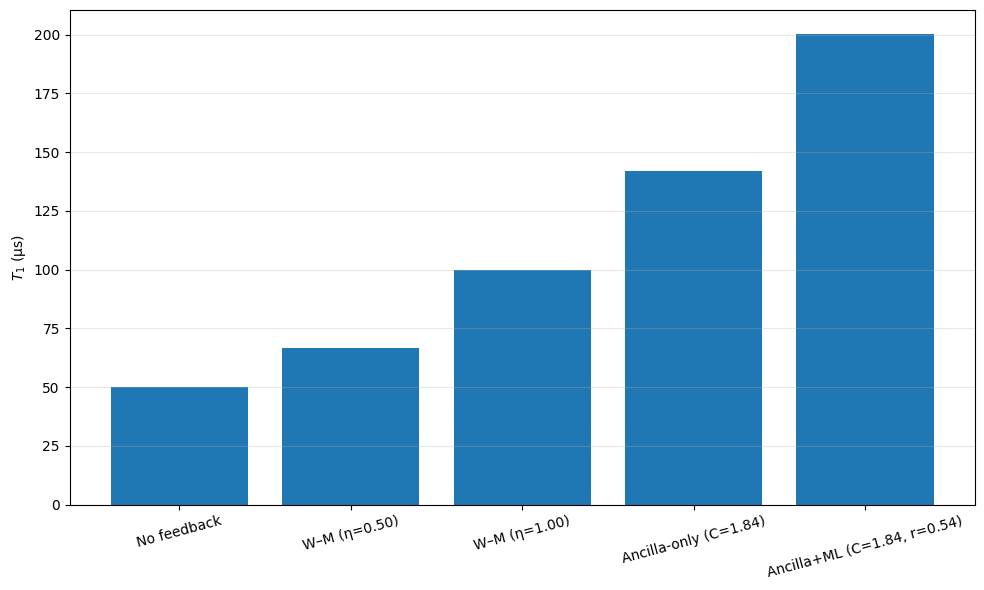}
    \caption{Effective relaxation times \(T_1\) for the five feedback-control configurations.
    The baseline (no feedback) corresponds to \(T_1 = 50~\mu\text{s}\);
    W--M feedback improves lifetime by up to \(2\times\);
    ancilla-assisted control extends it to \(142~\mu\text{s}\);
    and the ML-assisted configuration reaches \(T_1 \approx 201~\mu\text{s}\).}
    \label{fig:fig6}
\end{figure}

To assess the impact of the three feedback-control architectures, classical Wiseman-Milburn (W--M), ancilla-assisted, and ancilla+ML, we evaluate three complementary observables: the effective lifetime \(T_1\), the excited-state population \(P_e(t)\), and the integrated energy-retention function \(E(T)\). 
All quantities were computed using IBM-scale parameters: a bare lifetime \(T_1 = 50~\mu\text{s}\), detection efficiencies \(\eta \in \{0.50, 1.00\}\), cooperativity \(C = 1.84\), and correlation coefficient between the predicted and true currents \(r = 0.54\). 
The analytical relations for each case are summarized in Table~\ref{table2}.\\
Figure~\ref{fig:fig6} compares the effective lifetimes for all feedback schemes in a bar histogram.
Without feedback, the qubit lifetime equals the baseline \(T_1 = 50~\mu\text{s}\).
W-M feedback increases the lifetime depending on the detection efficiency, yielding \(T_1 = 66.7~\mu\text{s}\) for \(\eta=0.5\) and \(T_1 = 100~\mu\text{s}\) for \(\eta=1.0\), corresponding to 1.3× and 2× improvement, respectively.
The ancilla-assisted feedback further extends the lifetime to \(142~\mu\text{s}\) (\(C=1.84\)), while the ML-assisted variant achieves \(T_1 = 201~\mu\text{s}\), approximately a fourfold enhancement.
The monotonic increase seen in Fig.~\ref{fig:fig5} confirms that adding coherence and prediction progressively improves the system’s resistance to relaxation.

\begin{figure}[t]
    \centering
    \includegraphics[width=1.0\linewidth]{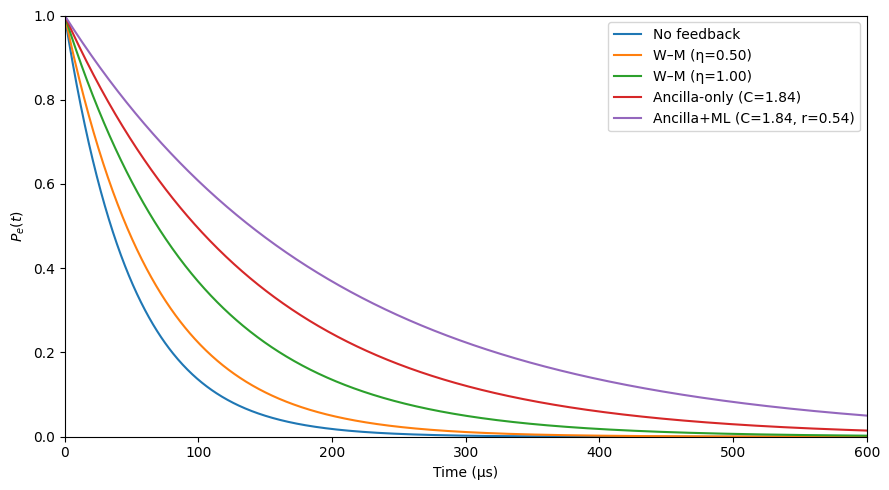}
    \caption{Time evolution of the excited-state population \(P_e(t)\) under different feedback schemes.
    The decay slows progressively from the uncontrolled case to W--M, ancilla-only, and finally ancilla+ML, indicating successive suppression of the effective decay rate.}
    \label{fig:fig7}
\end{figure}

Figure~\ref{fig:fig7} presents the time evolution of the excited-state population.
Each curve exhibits a single exponential decay characterized by the effective rate in Table~\ref{table2}.
The uncontrolled system decays most rapidly, followed sequentially by W--M feedback (\(\eta=0.5\) and \(\eta=1.0\)), the ancilla-only case, and finally the ML-enhanced feedback, which shows the slowest decay.
At \(t=100~\mu\text{s}\), the W-M (\(\eta=1\)) population has dropped to about 37\%, whereas the ancilla+ML configuration retains approximately 61\% of the initial excitation.
This clear separation illustrates the predictive scheme’s ability to delay relaxation by compensating for measurement delay and environmental fluctuations.
\begin{figure}[t]
    \centering    \includegraphics[width=1.0\linewidth]{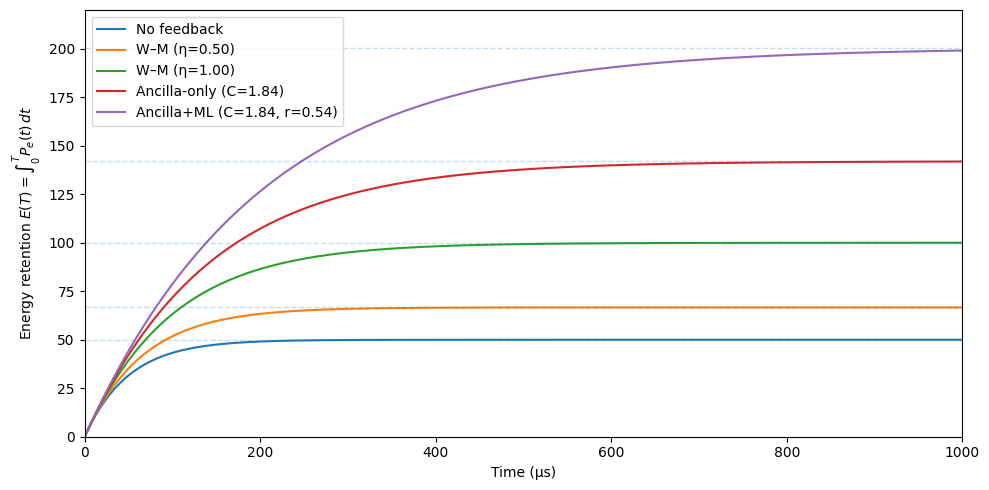}
    \caption{Energy-retention integral \(E(T)\) versus time for different feedback schemes.
    Each curve saturates at its effective lifetime \(T_1^{(\mathrm{eff})}\), validating the lifetime hierarchy observed in Fig.~\ref{fig:fig6}. 
    The ancilla+ML configuration exhibits both the highest plateau and the slowest saturation.}
    \label{fig:fig8}
\end{figure}

Figure~\ref{fig:fig8} plots the \emph{energy-retention} curve, defined as the
area under the excited-state population up to time $T$.
Operationally, it is the \emph{cumulative time} the qubit spends excited between $0$ and $T$.
Because the qubit’s internal energy is proportional to its excited-state population, this area is directly proportional to the \emph{total stored energy over the interval}.
Equivalently, it quantifies how much excited-state “budget” the controller preserves before that energy is finally emitted into the environment. At short times the curves rise nearly linearly, reflecting that the qubit has not yet had time to relax and the controller is still “banking” energy.
As time grows, each curve bends over and approaches a plateau: once almost all excitation has been emitted, accumulating additional “stored energy over time” becomes negligible, so the curve saturates.
The height of this plateau is \emph{exactly} the effective lifetime reported in Fig.~\ref{fig:fig6},
so the energy-retention plot provides an integral check of the lifetime analysis: higher $T_1^{\mathrm{(eff)}}$ means higher plateau. The saturation level tells us the \emph{total} usable energy preserved by a given control strategy before dissipation wins.
Faster saturation means the excited state is depleted quickly (little energy preserved); delayed saturation means the controller keeps the qubit excited for longer, accumulating more stored energy.
In our data, no feedback and classical W-M control saturate within $\sim\!200~\mu$s, indicating that most of the excitation has been lost by then.
Ancilla-assisted control pushes the plateau higher and delays saturation, showing that coherent exchange recycles part of the would-be emission back into the system.
With ML prediction, the curve climbs highest and saturates slowest: the controller anticipates the outgoing field, applies a phase-aligned correction, and thereby \emph{maximizes} the time-integrated stored energy. Energy-retention makes the comparison intuitive: it condenses the entire $P_e(t)$ trajectory into one scalar “area under the curve.”
Across all schemes the ordering is consistent with Fig.~\ref{fig:fig5}:
no feedback $<$ W-M $<$ ancilla $<$ ancilla+ML.
In short, strategies that extend the lifetime also deliver the largest retained-energy plateau, and the predictive (ancilla+ML) loop is the most effective at both \emph{slowing} decay and \emph{banking} energy over time.\\
\revred{The predictor is intentionally lightweight (small MLP, $W=5$)
and can be evaluated in microseconds on standard FPGA or CPU hardware.
In multi-qubit devices, the natural deployment is local: each qubit or
module runs an independent predictor using its own measurement record,
leading to approximately linear scaling. Exploring global predictors
for correlated records is an interesting direction for future work.}
\section{Conclusion}
\label{sec:sec5}

This work presents a coherent pathway to suppress amplitude damping by progressively enriching a measurement–feedback loop with quantum coherence and predictive inference. We benchmarked three architectures, classical Wiseman-Milburn (W-M) feedback, coherent ancilla–assisted feedback, and ancilla feedback augmented with a short-horizon machine-learning (ML) predictor, against natural decay, and evaluated them via effective lifetime, excited-state population decay, and time-integrated energy retention. The closed-form relations used for these comparisons are compiled in Table~\ref{table2}.

At the baseline, with no feedback, the qubit relaxes at its natural rate: the lifetime equals the intrinsic \(T_1\), the population decays accordingly, and the energy-retention integral saturates at \(T_1\). Classical W-M feedback improves upon this baseline in direct proportion to measurement efficiency, slowing population decay and raising the energy-retention plateau up to the efficiency-limited lifetime. Introducing a coherently coupled ancilla then overcomes the single-quadrature limitation of classical control: by opening a Hamiltonian exchange channel, it reduces the effective emission rate even without changes to detection, yielding a lifetime increase set by the cooperativity and a commensurate rise in the retained energy. Finally, adding an ML predictor aligns the corrective drive with the \emph{future} emitted field and compensates loop delay, producing the strongest overall improvement: the longest lifetime, the slowest decay, and the highest energy-retention plateau.\\
These outcomes are mutually consistent: lifetime extension, slower population decay, and increased energy retention move in lockstep because each strategy achieves a further reduction of the effective decay rate. Conceptually, coherence (via the ancilla) recovers information lost in single-quadrature classical feedback, while prediction (via a lightweight supervised model) mitigates unavoidable hardware latency. Practically, the approach is incremental and compatible with standard cavity- and circuit-QED platforms: begin with W-M feedback, add an ancilla path to restore the missing quadrature, and incorporate a short-horizon predictor trained on the homodyne record to anticipate the outgoing field. The resulting gains longer effective lifetimes, reduced decay rates, and larger energy-retention plateaus translate directly into deeper algorithmic circuits and reduced error-correction overheads.

In summary, information that would otherwise be lost to the environment can be captured, processed, and coherently returned to the qubit with the correct phase and timing. In the parameter regimes considered here, classical W-M control improves upon natural decay, ancilla-assisted control provides a further and substantial gain, and the ancilla+ML architecture delivers the largest and most practically relevant enhancement (see Table~\ref{table2}).

\begin{table*}[!t]
\centering
\small
\renewcommand{\arraystretch}{1.2}
\setlength{\tabcolsep}{8pt}
\caption{Comparison of feedback schemes for amplitude-damping suppression. 
Here $P_e(t)$ is the excited-state population, $\Gamma_{\mathrm{eff}}$ the effective decay rate, and $T_1^{\mathrm{eff}} = 1/\Gamma_{\mathrm{eff}}$ the corresponding lifetime.}
\label{table2}
\begin{tabular}{lccc}
\hline
\textbf{Scheme} 
& $\mathbf{P_e(t)}$ 
& $\mathbf{\Gamma_{\mathrm{eff}}}$ 
& $\mathbf{T_1^{\mathrm{eff}}}$ \\
\hline
No feedback 
& $P_e(0)\,e^{-\gamma t}$
& $\gamma$
& $1/\gamma$ \\

W--M (optimal, efficiency $\eta$)
& $P_e(0)\,e^{-\gamma(1-\eta/2)t}$
& $\gamma(1-\eta/2)$
& $1/\!\bigl[\gamma(1-\eta/2)\bigr]$ \\

Ancilla-only (cooperativity $C = 4g^2/(\kappa\gamma)$)
& $P_e(0)\,e^{-\frac{\gamma}{1+C}t}$
& $\gamma/(1+C)$
& $(1+C)/\gamma$ \\

Ancilla + ML (prediction quality $r = \mathrm{corr}(\widehat{I}, I)$)
& $P_e(0)\,e^{-\frac{\gamma}{1+C}(1-r^2)t}$
& $\dfrac{\gamma}{1+C}(1-r^2)$
& $\dfrac{1+C}{\gamma(1-r^2)}$ \\
\hline
\end{tabular}
\end{table*}

\bibliographystyle{IEEEtran}
\bibliography{main}
\appendices

\section{Derivation of Eq.~(\ref{eq:eq17})}
\label{appendixa}

Starting from the master equation with spontaneous emission at rate \(\gamma\),
\begin{equation}
\dot\rho(t)=\gamma\,\mathcal{D}[\sigma_-]\rho(t)
=\gamma\!\left(\sigma_- \rho \sigma_+ - \tfrac{1}{2}\{\sigma_+\sigma_-,\rho\}\right),
\end{equation}
where the Lindblad dissipator is defined as
\begin{equation}
\mathcal{D}[L]\rho = L\rho L^\dagger - \tfrac{1}{2}\{L^\dagger L, \rho\}, 
\end{equation}
 where $\{X,Y\}=XY+YX$ denotes the anticommutator. This equation describes the irreversible loss of excitation from the upper state
\(|e\rangle\) to the ground state \(|g\rangle\).

The excited-state population is defined by
\begin{equation}
P_e(t)=\operatorname{Tr}\!\big[\rho(t)\,\Pi_e\big],
\qquad 
\Pi_e=\lvert e\rangle\langle e\rvert.
\end{equation}
Differentiating and using linearity of the trace gives
\begin{equation}
\dot P_e(t)=\operatorname{Tr}\!\big[\dot\rho(t)\,\Pi_e\big]
= \gamma\,\operatorname{Tr}\!\big[\mathcal{D}[\sigma_-]\rho(t)\,\Pi_e\big].
\end{equation}
Expanding the dissipator and applying the cyclic property
\(\operatorname{Tr}[AB]=\operatorname{Tr}[BA]\),
\begin{align}
\operatorname{Tr}\!\big[\mathcal{D}[\sigma_-]\rho\,\Pi_e\big]
&= \operatorname{Tr}\!\big[\sigma_- \rho \sigma_+ \Pi_e\big]
- \tfrac{1}{2}\operatorname{Tr}\!\big[(\sigma_+\sigma_-)\rho \Pi_e\big] \nonumber \\
&\quad - \tfrac{1}{2}\operatorname{Tr}\!\big[\rho(\sigma_+\sigma_-)\Pi_e\big].
\end{align}
For a two-level atom
\(
\sigma_-=\lvert g\rangle\langle e\rvert,\;
\sigma_+=\lvert e\rangle\langle g\rvert,\;
\sigma_+\sigma_-=\Pi_e
\),
and
\(
\Pi_e\sigma_- = 0,\;
\sigma_+\Pi_e = 0.
\)
Hence the first term vanishes,
\(\operatorname{Tr}[\sigma_- \rho \sigma_+ \Pi_e]=0\),
and the remaining two yield
\begin{equation}
-\tfrac{1}{2}\operatorname{Tr}[\Pi_e \rho \Pi_e]
-\tfrac{1}{2}\operatorname{Tr}[\rho\,\Pi_e]
= -\,\operatorname{Tr}[\rho\,\Pi_e]
= -\,P_e(t),
\end{equation}
where \(\Pi_e^2=\Pi_e\) has been used.
Therefore,
\begin{equation}
\dot P_e(t) = -\,\gamma\,P_e(t), \qquad
P_e(t)=P_e(0)e^{-\gamma t},
\end{equation}
so that the effective decay rate is \(\Gamma_{\mathrm{eff}}=\gamma\) and the lifetime
\(T_1=\gamma^{-1}\).

\section{Derivation of the Wiseman--Milburn Feedback Equation}
\label{appendixb}

Because the excited-state probability is related to the population-difference
operator by
\begin{equation}
P_e = \frac{1}{2}\big(1+\langle\sigma_z\rangle\big),
\label{eq:Pe-sigmaz-relation}
\end{equation}
its time derivative is directly determined by the evolution of
$\langle\sigma_z\rangle$.  In this appendix we show that feedback modifies
the decay to
\begin{equation}
\dot P_e = -\,\Gamma(\lambda,\eta)\,P_e,
\qquad
P_e(t)=P_e(0)e^{-\Gamma(\lambda,\eta)t},
\label{eq:Pe-feedback-solution}
\end{equation}
with an effective rate $\Gamma(\lambda,\eta)$ that depends on the feedback gain
$\lambda$ and the detection efficiency $\eta$.
To obtain this result we first derive $d\langle\sigma_z\rangle/dt$ from the
Wiseman-Milburn master equation.

In the W-M feedback scheme, the detector measures the light emitted by
the atom and immediately applies a feedback signal back to the atom.  
If the homodyne detector is not perfect, only a fraction of the emitted light is detected.
We describe the detector efficiency by a number $
0 \le \eta \le 1$, where $\eta=1$ represents perfect detection and $\eta=0$ means no detection at all.

The unconditional master equation for the system’s density matrix $\rho$ is
\begin{equation}
\dot{\rho}
= \mathcal{D}\!\big[c - i\sqrt{\eta}\,F\big]\rho
+ (1-\eta)\,\mathcal{D}[F]\rho,
\label{eq:wm-ineff}
\end{equation}
where, $c=\sqrt{\gamma}\,\sigma_-$, $
F=\lambda\,\sigma_y$,   $\gamma$ is the natural decay rate, $\lambda$ is the feedback strength,
$\sigma_- = |g\rangle\langle e|$ is the lowering operator, and
$\sigma_y$ is the Pauli \(y\) matrix.

In experiments we do not directly observe $\rho$; instead we measure
expectation values \(\langle A\rangle = \operatorname{Tr}(A\rho)\)
of physical observables \(A\).
To find their time evolution we differentiate:
\begin{equation}
\frac{d}{dt}\langle A\rangle
= \frac{d}{dt}\operatorname{Tr}(A\rho)
= \operatorname{Tr}\!\big(A\,\dot{\rho}\big).
\label{eq:dA-start}
\end{equation}
Substituting Eq.~\eqref{eq:wm-ineff} for $\dot\rho$ gives
\begin{equation}
\frac{d}{dt}\langle A\rangle
= \operatorname{Tr}\!\big(A\,\mathcal{D}[c-i\sqrt{\eta}F]\rho\big)
+ (1-\eta)\operatorname{Tr}\!\big(A\,\mathcal{D}[F]\rho\big).
\label{eq:dA-middle}
\end{equation}

It is often more convenient to let the dissipator act on \(A\) instead of on
\(\rho\).
For any operator \(L\) we therefore define the
\emph{adjoint dissipator}
\begin{equation}
\mathcal{D}^\dagger[L]A
= L^\dagger A L - \tfrac{1}{2}\{L^\dagger L, A\},
\end{equation}
which satisfies the trace identity
\begin{equation}
\operatorname{Tr}\!\big(A\,\mathcal{D}[L]\rho\big)
= \operatorname{Tr}\!\big(\mathcal{D}^\dagger[L]A\,\rho\big).
\label{eq:D-adjoint-id}
\end{equation}
Using this in Eq.~\eqref{eq:dA-middle} gives the compact and general formula
\begin{equation}
\frac{d}{dt}\langle A\rangle
= \operatorname{Tr}\!\big[\mathcal{D}^\dagger[c-i\sqrt{\eta}F]A\,\rho\big]
+ (1-\eta)\operatorname{Tr}\!\big[\mathcal{D}^\dagger[F]A\,\rho\big].
\label{eq:dA-final}
\end{equation}
This relation is crucial: it connects the master equation for $\rho$ with
the measurable time evolution of any observable $A$.

To study population dynamics we now set \(A=\sigma_z\), the population-difference
operator, and use the Pauli-matrix relations
\[
\begin{aligned}
&\sigma_\pm = \tfrac{1}{2}(\sigma_x \pm i\sigma_y), \qquad
\sigma_+\sigma_- = \tfrac{1}{2}(I+\sigma_z), \\
&\sigma_-\sigma_+ = \tfrac{1}{2}(I-\sigma_z), \qquad
\sigma_y^2 = I, \\
&\sigma_y\sigma_z = -\,\sigma_z\sigma_y = 2i(\sigma_+ - \sigma_-).
\end{aligned}
\]

After substituting \(c = \sqrt{\gamma}\,\sigma_-\) and \(F = \lambda\sigma_y\),
and expanding the commutators step by step, one finds
\begin{align}
\mathcal{D}^\dagger[c-i\sqrt{\eta}F]\sigma_z
&= -\big(\gamma - 2\sqrt{\eta\gamma}\,\lambda + 2\lambda^2\big)\,(\sigma_z + I),
\\[3pt]
\mathcal{D}^\dagger[F]\sigma_z
&= -2\lambda^2\,\sigma_z.
\end{align}
Substituting these into Eq.~\eqref{eq:dA-final} and simplifying yields
\begin{equation}
\frac{d}{dt}\langle\sigma_z\rangle
= -\,\Gamma(\lambda,\eta)\,\big(\langle\sigma_z\rangle+1\big),
\label{eq:Gamma-eta}
\end{equation}

where $\Gamma(\lambda,\eta)=\gamma - 2\sqrt{\eta\gamma}\,\lambda + 2\lambda^2$. Combining Eq.~\eqref{eq:Gamma-eta} with the relation
\eqref{eq:Pe-sigmaz-relation} immediately gives
\[
\dot P_e = -\,\Gamma(\lambda,\eta)\,P_e,
\qquad
P_e(t)=P_e(0)e^{-\Gamma(\lambda,\eta)t},
\]
which is Eq.~\eqref{eq:Pe-feedback-solution}.

Minimizing \(\Gamma(\lambda,\eta)\) with respect to $\lambda$ gives the optimal
feedback strength
\[
\frac{d\Gamma}{d\lambda}
= -2\sqrt{\eta\gamma} + 4\lambda = 0
\quad\Rightarrow\quad
\lambda^\star = \tfrac{1}{2}\sqrt{\eta\,\gamma},
\]
and since \(d^2\Gamma/d\lambda^2=4>0\) this is indeed a minimum.
At this optimum,
\begin{align}
  \Gamma_{\mathrm{WM}}
&= \Gamma(\lambda^\star,\eta)
= \gamma\!\left(1-\frac{\eta}{2}\right),\\
T_1^{\mathrm{WM}}
&= \frac{1}{\Gamma_{\mathrm{WM}}}
= \frac{1}{\gamma(1-\eta/2)}.  
\end{align}

For perfect detection (\(\eta=1\)),
\(\Gamma_{\mathrm{WM}}=\gamma/2\) and \(T_1^{\mathrm{WM}}=2/\gamma\),
so the feedback doubles the lifetime of the excited state.
For inefficient detection (\(\eta<1\)), the improvement is smaller because
less information is available to the feedback loop.

W-M feedback can slow down spontaneous emission by an amount
that depends directly on the detection efficiency~$\eta$.
The better the detection, the more the feedback can suppress decay,
and in the ideal limit of $\eta=1$, the lifetime doubles compared with
natural spontaneous emission.

\section{Derivation of the Ancilla and ML Feedback Rates}
\label{appendixc}

This Appendix presents a detailed derivation of the effective decay rates
$\Gamma_{\mathrm{anc}}$ and $\Gamma_{\mathrm{ML}}$ corresponding to the
ancilla–assisted and machine–learning–enhanced feedback schemes.
The goal is to clarify how coupling a fast–decaying ancilla qubit to a
more slowly relaxing system qubit, together with predictive feedback,
can suppress the relaxation of the system.

\subsection*{A. Joint Master Equation}

We consider a system qubit $S$ coupled to an ancilla qubit $A$
with joint density matrix $\rho_{SA}$.
Their dynamics are governed by the Lindblad master equation

\begin{align}
\dot{\rho}_{SA}
&=
-\frac{i}{\hbar}[H_S+H_A+H_{\mathrm{int}},\,\rho_{SA}]
+ \gamma\,\mathcal{D}[\sigma_-^{(S)}]\rho_{SA}
\nonumber\\
&\quad
+ \kappa\,\mathcal{D}[\sigma_-^{(A)}]\rho_{SA},
\label{eq:C1}
\end{align}

where the Lindblad dissipator is

\[
\mathcal{D}[L]\rho
=
L\rho L^\dagger
-\frac12(L^\dagger L\rho+\rho L^\dagger L).
\]

Here

\begin{itemize}
\item $\gamma$ is the spontaneous emission rate of the system qubit,
\item $\kappa$ is the relaxation rate of the ancilla,
\item $g$ is the coherent coupling strength.
\end{itemize}

The two qubits exchange excitations through

\begin{equation}
H_{\mathrm{int}}
=
\hbar g
(\sigma_+^{(S)}\sigma_-^{(A)}
+
\sigma_-^{(S)}\sigma_+^{(A)}),
\label{eq:C2}
\end{equation}

which enables coherent excitation transfer between the system and the ancilla.

\subsection*{B. Fast–Ancilla Regime}

We assume the ancilla relaxes much faster than the system,

\[
\kappa \gg g, \gamma .
\]

In this regime the ancilla rapidly reaches a quasi–steady state and follows
the system dynamics almost instantaneously. This separation of timescales
allows us to eliminate the ancilla variables adiabatically.

\subsection*{C. Adiabatic Elimination of the Ancilla}

We first determine how the ancilla responds to the system dynamics.

The system qubit ($S$) is coherently coupled to the ancilla qubit ($A$)
with interaction strength $g$, while the ancilla decays rapidly to the
environment with rate $\kappa$.

The lowering operators are

\begin{itemize}
\item $\sigma_-^{(S)}$ : lowering operator of the system qubit
\item $\sigma_-^{(A)}$ : lowering operator of the ancilla qubit
\end{itemize}

The expectation value of the ancilla coherence evolves according to

\begin{equation}
\frac{d}{dt}\langle\sigma_-^{(A)}\rangle
=
-\frac{\kappa}{2}\langle\sigma_-^{(A)}\rangle
- i g\langle\sigma_-^{(S)}\sigma_z^{(A)}\rangle .
\end{equation}

Here

\begin{itemize}
\item $g$ is the coherent system–ancilla coupling strength,
\item $\kappa$ is the decay rate of the ancilla,
\item $\sigma_z^{(A)}$ is the Pauli-$z$ operator of the ancilla.
\end{itemize}

\paragraph{Ground-state approximation}

Because the ancilla relaxes much faster than the system ($\kappa \gg g$),
it remains close to its ground state. Therefore

\[
\langle\sigma_z^{(A)}\rangle \approx -1 .
\]

Substituting this approximation gives

\[
\frac{d}{dt}\langle\sigma_-^{(A)}\rangle
\approx
-\frac{\kappa}{2}\langle\sigma_-^{(A)}\rangle
+ i g\langle\sigma_-^{(S)}\rangle .
\]

\paragraph{Fast relaxation}

Since the ancilla relaxes very quickly, its dynamics reach steady state
almost instantaneously. We therefore set

\[
\dot{\langle\sigma_-^{(A)}\rangle}\approx0 .
\]

Solving for the steady-state value yields

\begin{equation}
\langle\sigma_-^{(A)}\rangle
\approx
\frac{2ig}{\kappa}\langle\sigma_-^{(S)}\rangle .
\end{equation}

Thus the ancilla coherence simply follows the system coherence.

At the operator level we therefore write

\[
\sigma_-^{(A)}
\simeq
\frac{2ig}{\kappa}\sigma_-^{(S)} .
\]

This result shows that the ancilla responds almost instantaneously
to the system excitation.

\subsection*{D. Effective Dissipation Channel}

We now substitute the above relation into the dissipative term
associated with the ancilla decay.

The ancilla dissipator is

\[
\kappa \mathcal{D}[\sigma_-^{(A)}]\rho ,
\]

where the Lindblad dissipator is defined as

\[
\mathcal{D}[L]\rho
=
L\rho L^\dagger
-
\frac12
\left(
L^\dagger L\rho
+
\rho L^\dagger L
\right).
\]

Using the relation
\[
\sigma_-^{(A)} \simeq \frac{2ig}{\kappa}\sigma_-^{(S)},
\]

we obtain

\[
\kappa\mathcal{D}[\sigma_-^{(A)}]\rho
\approx
\frac{4g^2}{\kappa}
\mathcal{D}[\sigma_-^{(S)}]\rho .
\]

Tracing out the ancilla degrees of freedom gives the effective
system master equation

\begin{equation}
\dot{\rho}_S
=
\left(
\gamma + \frac{4g^2}{\kappa}
\right)
\mathcal{D}[\sigma_-^{(S)}]\rho_S .
\end{equation}

Here

\begin{itemize}
\item $\gamma$ is the intrinsic decay rate of the system qubit
\item $\frac{4g^2}{\kappa}$ represents the additional decay channel
introduced by the ancilla.
\end{itemize}

This intermediate result corresponds to a purely dissipative coupling
between the system and the ancilla.

\subsection*{E. Cooperativity and Coherent Feedback}

In the coherent–feedback configuration considered in this work,
the ancilla does not simply introduce an additional loss channel.
Instead, the ancilla is driven by the measurement-based feedback signal
and remains coherently coupled to the system qubit.

As a result, the radiation emitted through the ancilla forms a
secondary emission pathway that interferes with the direct
spontaneous emission of the system.

\paragraph{Feedback-induced emission channel}

From the adiabatic elimination derived in the previous subsection,
the ancilla-mediated decay channel produces an effective rate

\begin{equation}
\gamma_{\mathrm{fb}}
=
\frac{4g^2}{\kappa},
\end{equation}

where

\begin{itemize}
\item $g$ is the coherent coupling strength between the system and the ancilla,
\item $\kappa$ is the decay rate of the ancilla.
\end{itemize}

Physically, $\gamma_{\mathrm{fb}}$ represents the emission rate
associated with the ancilla-mediated feedback pathway.

\paragraph{Interference of emission amplitudes}

Because the feedback loop is coherent, the field emitted through
the ancilla interferes with the system’s direct emission.
The effective collapse operator describing the net emission
amplitude therefore becomes

\begin{equation}
c_{\mathrm{eff}}
=
\left(
\sqrt{\gamma}
-
\sqrt{\gamma_{\mathrm{fb}}}
\right)
\sigma_-^{(S)} ,
\end{equation}

where $\gamma$ is the intrinsic spontaneous-emission rate of the system.

\paragraph{Cooperativity}

To express this result in dimensionless form,
it is convenient to introduce the cooperativity parameter

\begin{equation}
C=\frac{4g^2}{\kappa\gamma}.
\end{equation}

The cooperativity compares the strength of coherent
system–ancilla coupling ($g$) with the dissipative rates
$\kappa$ and $\gamma$.

A larger value of $C$ means that excitation can cycle
multiple times between the system and the ancilla before
being lost to the environment.

\paragraph{Effective decay rate}

Using the definition of $C$, the effective decay rate of the
system becomes

\begin{equation}
\Gamma_{\mathrm{anc}}
=
\frac{\gamma}{1+C}.
\end{equation}

\paragraph{Lifetime enhancement}

The corresponding relaxation time is the inverse of the decay rate,

\begin{equation}
T_1^{(\mathrm{anc})}
=
\Gamma_{\mathrm{anc}}^{-1}
=
\frac{1+C}{\gamma}.
\end{equation}

Thus the ancilla-assisted coherent feedback suppresses
the spontaneous emission rate by the factor $1/(1+C)$
and increases the lifetime of the system by the factor $(1+C)$.
\subsection*{F. Inclusion of ML Predictive Feedback}

Finally we incorporate the machine-learning prediction that
compensates the feedback delay.

The ML model predicts the future homodyne current

\[
\widehat{I}(t+\tau),
\]

which drives the control signal applied to the ancilla

\[
u_{\mathrm{ML}}(t)=\lambda\,\widehat{I}(t+\tau).
\]

The prediction accuracy is quantified using the correlation coefficient

\begin{equation}
r=
\mathrm{corr}(\widehat{I}(t+\tau), I(t+\tau)),
\qquad 0\le r\le1.
\end{equation}

\begin{itemize}
\item $r=1$ corresponds to perfect prediction
\item $r=0$ corresponds to no predictive information
\end{itemize}

The remaining uncorrelated noise scales as $(1-r^2)$.

This effectively rescales the system collapse operator

\[
\sigma_-^{(S)}
\rightarrow
(1-r^2)^{1/2}\sigma_-^{(S)} .
\]

Substituting this into the ancilla-assisted decay rate yields

\begin{equation}
\Gamma_{\mathrm{ML}}
=
\frac{\gamma}{1+C}(1-r^2).
\end{equation}

The corresponding relaxation time becomes

\begin{equation}
T_1^{(\mathrm{ML})}
=
\frac{1+C}{\gamma(1-r^2)}.
\end{equation}

Thus the machine-learning prediction further suppresses
the effective decay rate by reducing the feedback noise.
\end{document}